\documentclass[twocolumn,showpacs,prd]{revtex4}
\usepackage{amsmath}
\usepackage{mathrsfs,bm,multirow}
\usepackage{longtable,lscape}
\usepackage{txfonts}
\usepackage{amssymb}
\usepackage{indentfirst}
\usepackage{graphicx,,booktabs}
\usepackage{color}
\usepackage{amssymb}
\begin{document}

\title{Mass Spectra of $0^{+-}$, $1^{-+}$, and $2^{+-}$ Exotic Glueballs}
\author{Liang Tang$^{1,3}$}
\email{tangl@mail.hebtu.edu.cn}
\author{Cong-Feng Qiao$^{2,3}$\footnote{Corresponding author}}
\email{qiaocf@ucas.ac.cn}
\affiliation{$^1$Department of Physics, Hebei Normal University, Shijiazhuang 050024, China\\
$^2$School of Physics, University of Chinese Academy of Sciences - YuQuan Road 19A, Beijing 100049, China\\
$^3$CAS Center for Excellence in Particle Physics, Beijing 100049, China}

\begin{abstract}

  With appropriate interpolating currents the mass spectra of $0^{+-}$, $1^{-+}$, and $2^{+-}$ oddballs are studied in the framework of QCD sum rules (QCDSR). We find there exits one stable $0^{+-}$ oddball with mass of $4.57 \pm 0.13 \, \text{GeV}$, and one stable $2^{+-}$ oddball with mass of $6.06 \pm 0.13 \, \text{GeV}$, whereas, no stable $1^{-+}$ oddball shows up. The possible production and decay modes of these glueballs with unconventional quantum numbers are analyzed, which are hopefully measurable in either BELLEII, PANDA, Super-B or LHCb experiments.

\end{abstract}
 \maketitle

\section{Introduction}

Quantum chromodynamics (QCD) is the underlying theory of hadronic interaction. In the high energy regime, it has been tested up to the 1\% level due to asymptotic freedom \cite{Gross:1973id}. However, the nonperturbative aspect related to the hadron spectrum is difficult to be calculated from first principles because of the confinement \cite{Wilson:1974sk}. A unique attempt in understanding the nonperturbative aspect of QCD is to study the glueball ($gg$, $ggg$, $\cdots$), where the gauge field plays a more important dynamical role than in ordinary hadrons. This has created much interest in theory and experiment for quite a long time.

In the literature, many theoretical investigations on glueball were made through various techniques, including lattice QCD \cite{Morningstar:1999rf, Mathieu:2008me, Chen:2005mg, Gregory:2012hu, Ishikawa:1982kk}, the flux tube model \cite{Isgur:1984bm}, the MIT bag model \cite{Chodos:1974je, Jaffe:1975fd}, the Coulomb gauge model \cite{Szczepaniak:1995cw}, the holographic model \cite{Colangelo:2007pt, Bellantuono:2015fia, Brunner:2015yha, Brunner:2015oqa}, and QCD sum rules (QCDSR) \cite{Shifman, Reinders:1984sr, twogluon0++, Huang:1998wj, Narison:1996fm, twogluon0-+, Latorre:1987wt, Lu:1996tp, Hao:2005hu}. Of these techniques, the QCDSR, developed more than 30 years ago by Shifman, Vainshtein, and Zakharov (SVZ) \cite{Shifman}, has some peculiar advantages in the study of hadron phenomenology. Its starting point in evaluating the properties of the ground-state hadron is to construct the current, which possesses the foremost information about the concerned hadron, like quantum numbers and the constituent quark or gluon. By using the current, one can then construct the two-point correlation function, which has two representations: the QCD representation and the phenomenological representation. Equating these two representations, the QCDSR will be formally established.

In the framework of QCDSR, the two-gluon glueballs with conventional quantum numbers of $0^{++}$ \cite{twogluon0++, Huang:1998wj, Narison:1996fm}  and $0^{-+}$ \cite{Narison:1996fm, twogluon0-+} have been studied extensively. Note that even the trigluon components of these glueballs were considered \cite{Latorre:1987wt, Lu:1996tp, Hao:2005hu}, which is enlightening for the research of this work.

Although the glueball has been searched for for many years in experiments, so far there has been no definite conclusion about it, mainly due to the mixing effect between glueballs and quark states, and lack of the knowledge about glueball production scheme and decay properties. Of these difficulties, from the experimental point of view, the most outstanding obstacle is how to disentangle the glueball from the mixed quarkonium states ( $q \bar{q}$ ). Fortunately, there is a class of glueballs, the unconventional glueballs, which with quantum numbers unaccessible by quark-antiquark bound states can avoid such problems. The quantum numbers of those glueballs include $J^{PC} = 0^{--}$, $0^{+-}$, $1^{-+}$, $2^{+-}$, $3^{-+}$, and so on. Note, according to $C$-parity conservation, glueballs with negative $C$ parity cannot be reached by two gluons, but have to be composed of at least three gluons. It should be noted that the $1^{-+}$ glueball also have to be made of at least three gluons, since the coupling of two transverse particles forbids the existence of $J=1$ states. This fact is known as Yang's theorem \cite{Yang:1950rg}. In the literature the term oddball has been used to describe glueballs having unconventional quantum numbers \cite{Matheus:2006xi} as well as 3 gluon glueballs with odd $J$, $P$, $C$ having conventional quantum numbers \cite{Szczepaniak:1995cw}. In this paper, we adopt the definition of oddball in \cite{Qiao:2014vva} to unify and avoid confusion.

Among various oddballs, special attention ought be paid to the $0^{--}$ ones, since they possess the lowest spin and their quantum number enables their production in the decays of vector quarkonium or quarkoniumlike states relatively easier. Ref. \cite{Qiao:2014vva} studied the $0^{--}$ oddballs via QCD Sum Rules, and found there exit two stable $0^{--}$ oddballs with masses of $3.81 \pm 0.12 \, \text{GeV}$ and $4.33 \pm 0.13 \, \text{GeV}$. The aim of this paper is to evaluate the other unconventional oddballs which have to be composed of at least three gluons (i.e., $J^{PC} = 0^{+-}$, $1^{-+}$, and $2^{+-}$) and discuss the feasibility of finding them in experiment.

This paper is organized in five sections. After the Introduction, in Sec.II we brief the method of QCD Sum Rules and construct the appropriate interpolating currents for oddballs. Sec.III gives the analytical results and numerical analyses for each oddball. In Sec.IV,  the possible production and decay modes of oddballs are investigated. The last section is left for discussion and conclusion.

\section{Formalism}

In order to calculate the mass spectra of the $0^{+-}$, $1^{-+}$, and $2^{+-}$ oddballs, one has to construct the appropriate currents for them. In practice a number of currents satisfy each the unconventional quantum numbers. However, after imposing the constraints of gauge invariance, Lorentz invariance, and $SU_c(3)$ symmetry, only a limited number of currents remain for each quantum number. The interpolating currents of the $0^{+-}$ oddballs are
\begin{eqnarray}
j^{0^{+-}, \; A}(x) & \!\!\! = \!\!\! & g_s^3 d^{a b c} [g^t_{\alpha \beta}(\partial) G^a_{\mu \nu}(x)][\partial_\alpha \partial_\beta G^b_{\nu \rho}(x)][G^c_{\rho \mu}(x)]\, , \label{current-0+-A}\\
j^{0^{+-}, \; B}(x) & \!\!\! = \!\!\! & g_s^3 d^{a b c} [g^t_{\alpha \beta}(\partial) G^a_{\mu \nu}(x)][\partial_\alpha \partial_\beta \tilde{G}^b_{\nu \rho}(x)][\tilde{G}^c_{\rho \mu}(x)]\, , \label{current-0+-B}\\
j^{0^{+-}, \; C}(x) & \!\!\! = \!\!\! & g_s^3 d^{a b c} [g^t_{\alpha \beta}(\partial) \tilde{G}^a_{\mu \nu}(x)][\partial_\alpha \partial_\beta G^b_{\nu \rho}(x)][\tilde{G}^c_{\rho \mu}(x)]\, ,\label{current-0+-C}\\
j^{0^{+-}, \; D}(x) & \!\!\! = \!\!\! & g_s^3 d^{a b c} [g^t_{\alpha \beta}(\partial) \tilde{G}^a_{\mu \nu}(x)][\partial_\alpha \partial_\beta \tilde{G}^b_{\nu \rho}(x)][G^c_{\rho \mu}(x)]\, ,\label{current-0+-D}
\end{eqnarray}
where $a$, $b$, and $c$ are color indices, $\mu$, $\nu$, $\rho$, $\alpha$, and $\beta$ are Lorentz indices, $d^{a b c}$ stands for the totally symmetric $SU_c(3)$ structure constant, $g^t_{\alpha \beta}(\partial)= g_{\alpha \beta}- \partial_\alpha \partial_\beta/\partial^2$, $G^a_{\mu \nu}$ denotes the gluon field strength tensor, and $\tilde{G}^a_{\mu \nu}$ is the dual gluon field strength tensor defined as $\tilde{G}^a_{\mu \nu} = \frac{1}{2} \epsilon_{\mu \nu \kappa \tau} G^a_{\kappa \tau}$\ .
Hereafter, for simplicity the four $0^{+-}$ currents in Eqs.(\ref{current-0+-A})-(\ref{current-0+-D}) will be referred as case $A$ to $D$, respectively, and they will be all taken into account in our analysis. These notations and conventions are suitable for the following currents with the other quantum numbers.

The interpolating currents of the $1^{-+}$ oddballs are
\begin{eqnarray}
j^{1^{-+}, \; A}_{\alpha}(x) & \!\! = \!\! & g_s^3 f^{a b c} \partial_\mu[G^a_{\mu \nu}(x)][G^b_{\nu \rho}(x)][G^c_{\rho \alpha}(x)]\, , \label{current-1-+A}\\
j^{1^{-+}, \; B}_{\alpha}(x) & \!\! = \!\! & g_s^3 f^{a b c} \partial_\mu[G^a_{\mu \nu}(x)][\tilde{G}^b_{\nu \rho}(x)][\tilde{G}^c_{\rho \alpha}(x)]\, , \label{current-1-+B}\\
j^{1^{-+}, \; C}_{\alpha}(x) & \!\! = \!\! & g_s^3 f^{a b c} \partial_\mu[\tilde{G}^a_{\mu \nu}(x)][G^b_{\nu \rho}(x)][\tilde{G}^c_{\rho \alpha}(x)]\, , \label{current-1-+C}\\
j^{1^{-+}, \; D}_{\alpha}(x) & \!\! = \!\! & g_s^3 f^{a b c} \partial_\mu[\tilde{G}^a_{\mu \nu}(x)][\tilde{G}^b_{\nu \rho}(x)][G^c_{\rho \alpha}(x)]\, , \label{current-1-+D}
\end{eqnarray}
where $f^{a b c}$ stands for the totally antisymmetric $SU_c(3)$ structure constant.

The interpolating currents of the $2^{+-}$ oddballs are
\begin{eqnarray}
j^{2^{+-}, \; A}_{\mu \alpha}(x) & \!\! = \!\! & g_s^3 d^{a b c} [G^a_{\mu \nu}(x)][G^b_{\nu \rho}(x)][G^c_{\rho \alpha}(x)]\, ,\label{current-2+-A}\\
j^{2^{+-}, \; B}_{\mu \alpha}(x) & \!\! = \!\! & g_s^3 d^{a b c} [G^a_{\mu \nu}(x)][\tilde{G}^b_{\nu \rho}(x)][\tilde{G}^c_{\rho \alpha}(x)]\, ,  \label{current-2+-B}\\
j^{2^{+-}, \; C}_{\mu \alpha}(x) & \!\! = \!\! & g_s^3 d^{a b c} [\tilde{G}^a_{\mu \nu}(x)][G^b_{\nu \rho}(x)][\tilde{G}^c_{\rho \alpha}(x)]\, ,  \label{current-2+-C}\\
j^{2^{+-}, \; D}_{\mu \alpha}(x) & \!\! = \!\! & g_s^3 d^{a b c} [\tilde{G}^a_{\mu \nu}(x)][\tilde{G}^b_{\nu \rho}(x)][G^c_{\rho \alpha}(x)]\, .\label{current-2+-D}
\end{eqnarray}

With currents of (\ref{current-0+-A})-(\ref{current-2+-D}), the two-point correlation functions can be readily established, i.e. ,
\begin{eqnarray}
\Pi^{J^{PC}, \; k}_{\alpha_1 \cdots \alpha_j, \; \beta_1 \cdots \beta_j}(q^2) \!\! = \!\! i \!\!\! \int \!\!\! d^4 \! x e^{i q \cdot x} \langle 0 | T \bigg\{ j^{J^{PC}, \; k}_{\alpha_1 \cdots \alpha_j}(x), j^{J^{PC}, \; k}_{\beta_1 \cdots \beta_j}(0)\bigg\}| 0 \rangle \; ,\label{correlation-1}
\end{eqnarray}
where the superscript $J^{PC}$ denotes the quantum number of the involved oddball, $k$ runs from $A$ to $D$, and $|0 \rangle$ denotes the physical vacuum. Here, the sets ($\alpha_1 \cdots \alpha_j$) and ($\beta_1 \cdots \beta_j$) respectively denote the Lorentz indices of the interpolating current that located at points x and 0, where the subscript $j$ represents the number of free Lorentz indices of the interpolating current.

Eq.(\ref{correlation-1}) has the following structure \cite{YBDai}
\begin{eqnarray}
& &i \int  d^4  x e^{i q \cdot x} \langle 0 | T \bigg\{ j^{J^{PC}, \; k}_{\alpha_1 \cdots \alpha_j}(x), j^{J^{PC}, \; k}_{\beta_1 \cdots \beta_j}(0)\bigg\}| 0 \rangle \nonumber \\
&=& T_{\alpha_1 \cdots \alpha_j, \; \beta_1 \cdots \beta_j}\Pi^k_{J^{PC}}(q^2) + \cdots \; ,
\end{eqnarray}
where ``$\cdots$" represents other structures which are independent of the correlation function $\Pi_{J^{PC}}^k(q^2)$. Here for the oddballs with $J=1$ and $2$, $T_{\alpha_1 \cdots \alpha_j, \; \beta_1 \cdots \beta_j}$ are of the form
\begin{eqnarray}
T_{\alpha_1, \; \beta_1} & = & g^t_{\alpha_1 \beta_1}(q) \; , \\
T_{\alpha_1 \alpha_2, \; \beta_1 \beta_2} & = & \frac{1}{2}\left[g^t_{\alpha_1 \beta_1}(q) + g^t_{\alpha_2 \beta_2}(q)\right] \nonumber \\ & - & \frac{1}{3} g^t_{\alpha_1 \alpha_2}(q) g^t_{\beta_1 \beta_2}(q) \; ,
\end{eqnarray}
with $g^t_{\alpha \beta}(q) = g_{\alpha \beta}- q_\alpha q_\beta / q^2$.

The QCD side of the correlation function can be obtained through the operator product expansion (OPE) and reads as
\begin{eqnarray}
\mathrm{\Pi_{J^{PC} }^{k, \; QCD}}(q^2) & \!\! = \!\! & a_0 (q^2)^{n} \ln\frac{- q^2}{\mu^2} + \left( b_0 + b_1 \ln \frac{-q^2}{\mu^2} \right)(q^2)^{n-2} \langle \alpha_s G^2 \rangle  \nonumber \\
& \!\!+ \!\! & \left( c_0 + c_1 \ln\frac{- q^2}{\mu^2} \right) (q^2)^{n-3} \langle g_s G^3 \rangle \nonumber \\
& \!\! + \!\! & d_0 (q^2)^{n-4} \langle \alpha_s G^2 \rangle^2 \ , \label{correlation-function-QCD}
\end{eqnarray}
where, $\langle \alpha_s G^2 \rangle$, $\langle g_s G^3\rangle$, and $\langle \alpha_s G^2 \rangle^2$ represent two-gluon, three-gluon, and four-gluon condensates, respectively; $\mu$ is the renormalization scale; and $n$ represents the corresponding power of $q^2$ for each oddball. For simplicity, we use $a_0$, $b_0$, $b_1$, $c_0$, $c_1$, and $d_0$ to represent the Wilson coefficients of operators with different dimensions in Eq.(\ref{correlation-function-QCD}).

On the phenomenological side, adopting the pole plus continuum parametrization of the hadronic spectral density, the imaginary part of the correlation function can be saturated as
\begin{eqnarray}
\frac{1}{\pi} \mathrm{Im\Pi_{J^{PC}}^{k, \; phe}}(s) & = & (f_{J^{PC}}^{k})^2 (M_{J^{PC}}^{k})^{2n} \delta \left(s - (M_{J^{PC}}^{k})^2 \right) \nonumber \\
& + & \rho_{J^{PC}}^{k} (s) \theta(s - s_0) \; .
\end{eqnarray}
Here $\rho_{J^{PC}}^{k}(s)$ is the spectral function of excited states and continuum states above the continuum threshold $\sqrt{s_0}$, $M^k_{J^{PC}}$ represents the mass of the $J^{PC}$ oddball, $f_{J^{PC}}^{k}$ stands for the coupling parameter.
Assuming $|G^{k}_{J^{PC}}>$ to be the oddball with the quantum number $J^{PC}$, the coupling parameter is defined by the following matrix element:
\begin{eqnarray}
  \langle 0 | j^{J^{PC}, \; k}_{\alpha_1 \cdots \alpha_j}| G^{k}_{J^{PC}} \rangle =f_{J^{PC}}^{k} \varepsilon_{\alpha_1 \cdots \alpha_j} \; ,
\end{eqnarray}
where $\varepsilon_{\alpha_1 \cdots \alpha_j}$ is the related polarization tensor.

Employing the dispersion relation on both QCD and phenomenological sides, i.e.,
\begin{eqnarray}
\Pi_{J^{PC}}^{k}(q^2) &\!\! = \!\!& \frac{1}{\pi} \int_0^\infty ds \frac{\text{Im} \Pi_{J^{PC}}^{k}(s)}{s - q^2} + \bigg(\Pi_{J^{PC}}^{k}(0) + q^2 \Pi_{J^{PC}}^{k \; \prime}(0) \nonumber \\
&\!\!\!\!+\!\!\!\!& \frac{1}{2} q^4 \Pi_{J^{PC}}^{k \; \prime \prime}(0) + \frac{1}{6} q^6 \Pi_{J^{PC}}^{k \; \prime \prime \prime}(0) \bigg) \; ,
\end{eqnarray}
where $\Pi_{J^{PC}}^{k}(0)$, $\Pi_{J^{PC}}^{k \; \prime}(0)$, $\Pi_{J^{PC}}^{k \; \prime \prime}(0)$, and $\Pi_J^{k \; \prime \prime \prime}(0)$ are constants relevant to the correlation function at the origin, then one can establish connection between QCD calculation (the QCD side) and the glueball properties (the phenomenological side),
\begin{eqnarray}
& & \frac{1}{\pi}  \int_0^{\infty}  \frac{\mathrm{Im\Pi_{J^{PC}}^{k, \; QCD}}(s)}{s - q^2} ds \nonumber \\ & = & \frac{(f_{J^{PC}}^{k})^2 (M_{J^{PC}}^{k})^{2n}}{(M_{J^{PC}}^{k})^2 - q^2}   + \int_{s_0}^{\infty} \!\! \frac{\rho_{J^{PC}}^{k}(s)\theta(s - s_0)}{s - q^2} ds \ .~~ \label{connection}
\end{eqnarray}

In order to take control of the contributions from higher order condensates in the OPE and the contributions from higher excited and continuum states on the phenomenological side, an effective and prevailing way is to perform the Borel transformation simultaneously on both sides of the QCDSR. That is
\begin{equation}
\hat{B}_{\tau}\equiv \lim_{-q^2\rightarrow \infty,n\rightarrow \infty
\atop
{-q^2\over n}=
{1\over\tau}}\frac{(q^2)^n}{(n-1)!}\left(- \frac{d}{dq^2}\right)^n\ ,
\end{equation}
where a parameter $\tau$, usually called the Borel parameter, is introduced.
After performing the Borel transformation, Eq.(\ref{connection}) then turns into
\begin{eqnarray}
& &\frac{1}{\pi} \int_0^{\infty}\! e^{-s\tau} \mathrm{Im\Pi_{J^{PC}}^{k, \; QCD}}(s) ds \nonumber \\
& = & (f^{k}_{J^{PC}})^2 (M^k_{J^{PC}})^{2n} e^{- \tau (M^k_{J^{PC}})^2} + \! \int_{s_0}^{\infty} \! \rho_{J^{PC}}^{k}(s) e^{- s \tau} ds\ .~~
\end{eqnarray}

Taking the quark-hadron duality approximation
\begin{equation}
\frac{1}{\pi}\int_{s_0}^{\infty}
e^{-s\tau}\mathrm{Im\Pi_{J^{PC}}^{k, \; QCD}}(s)ds \simeq
\int_{s_0}^{\infty} \rho^{k}_{J^{PC}}(s) e^{-s\tau} ds\; ,
\end{equation}
the moments $L^{k}_{J^{PC}, \; 0}$ and $L^{k}_{J^{PC}, \; 1}$ are achieved,
\begin{eqnarray}
L^{k}_{J^{PC}, \; 0}(\tau, s_0) & = & \frac{1}{\pi} \int_0^{s_0} e^{-s\tau} \mathrm{Im\Pi_{J^{PC}}^{k, \; QCD}}(s) ds \; , \label{R0} \\
L^{k}_{J^{PC}, \; 1}(\tau, s_0) & = & \frac{1}{\pi} \int_0^{s_0} s e^{-s\tau} \mathrm{Im\Pi_{J^{PC}}^{k, \; QCD}}(s) ds \; , \label{R1}
\end{eqnarray}
where $L^{k}_{J^{PC}, \; 1}(\tau, s_0)$ is obtained via $L^{k}_{J^{PC}, \; 1}(\tau, s_0)= - \partial L^{k}_{J^{PC}, \; 0}(\tau, s_0)/ \partial \tau$.
Then the $J^{PC}$ oddball mass is obtained in the form of the ratio of  $L^{k}_{J^{PC}, \; 1}(\tau, s_0)$ to $L^{k}_{J^{PC}, \; 0}(\tau, s_0)$, i.e. ,
\begin{eqnarray}
M^{k}_{J^{PC}}(\tau, s_0) = \sqrt{ \frac{L^k_{J^{PC}, \; 1} (\tau, s_0)}{L^k_{J^{PC}, \; 0} (\tau, s_0)}} \; , \label{mass}
\end{eqnarray}
where $k$ for cases $A$, $B$, $C$, and $D$.

\section{Analytical results and Numerical analyses}

After a lengthy calculation, the Wilson coefficients are obtained as follows. For the $0^{+-}$ oddballs, they are
\begin{eqnarray}
\begin{aligned}
& a_0^A \!\! = \!\! \frac{487}{143\times 2^6 \times 3^3} \frac{\alpha_s^3}{\pi} \; ,  & & b_0^A \!\! = \!\! \frac{5}{36}\pi \alpha_s^2 \;, & & b_1^A \!\! = \!\! 0 \; , \\
& c_0^A \!\! = \!\! - \frac{325}{72} \pi \alpha_s^3\;, & & c_1^A \!\! = \!\! - \frac{2125}{144} \pi \alpha_s^3 \;, & & d_0^A \!\! = \!\! 0 \; ; \\
& a_0^B \!\! = \!\! \frac{487}{143\times 2^6 \times 3^3} \frac{\alpha_s^3}{\pi} \; , & & b_0^B \!\! = \!\! \frac{5}{36}\pi \alpha_s^2 \;, & & b_1^B \!\! = \!\! 0 \; , \\
& c_0^B \!\! = \!\! \frac{7445}{144} \pi \alpha_s^3\;, & & c_1^B \!\! = \!\! \frac{1075}{96} \pi \alpha_s^3 \;, & & d_0^B \!\! = \!\! 0 \; ; \\
& a_0^C \!\! = \!\! \frac{487}{143\times 2^6 \times 3^3} \frac{\alpha_s^3}{\pi} \; , & & b_0^C \!\! = \!\! \frac{5}{36}\pi \alpha_s^2 \;, & & b_1^C \!\! = \!\! 0 \; , \\
& c_0^C \!\! = \!\! \frac{1955}{72} \pi \alpha_s^3\;, & & c_1^C \!\! = \!\! \frac{775}{144} \pi \alpha_s^3 \;, & & d_0^C \!\! = \!\! 0 \; ; \\
& a_0^D \!\! = \!\! \frac{487}{143\times 2^6 \times 3^3} \frac{\alpha_s^3}{\pi} \; , & & b_0^D \!\! = \!\! \frac{5}{36}\pi \alpha_s^2 \;, & & b_1^D \!\! = \!\! 0 \; , \\
& c_0^D \!\! = \!\!  \frac{235}{72} \pi \alpha_s^3\;, & & c_1^D \!\! = \!\! \frac{25}{32} \pi \alpha_s^3 \;, & & d_0^D \!\! = \!\! 0 \; ,
\end{aligned}
\end{eqnarray}
where we notice that except for $c^k_0$ and $c^k_1$, $a_0^k$, $b_0^k$, $b_1^k$, and $d_0^k$ are equal for case $A$ to $D$. This situation is similar to the $0^{--}$ oddballs in \cite{Qiao:2014vva}.

For the $1^{-+}$ oddballs, the Wilson coefficients are
\begin{eqnarray}
\begin{aligned}
& a_0^A \!\! = \!\! \frac{1}{1008} \frac{\alpha_s^3}{\pi} \; ,  & & b_0^A \!\! = \!\! -\frac{1}{72}\pi \alpha_s^2 \;, & & b_1^A \!\! = \!\! \frac{1}{12}\pi \alpha_s^2 \; , \\
& c_0^A \!\! = \!\! \frac{71}{96} \pi \alpha_s^3\;, & & c_1^A \!\! = \!\! \frac{23}{48} \pi \alpha_s^3 \;, & & d_0^A \!\! = \!\! \frac{1}{3}\pi^3 \alpha_s \; ; \\
& a_0^B \!\! = \!\! \frac{1}{1008 \pi} \frac{\alpha_s^3}{\pi} \; , & & b_0^B \!\! = \!\! \frac{23}{72}  \pi \alpha_s^2 \;, & & b_1^B \!\! = \!\! \frac{1}{12} \pi \alpha_s^2 \; , \\
& c_0^B \!\! = \!\! \frac{89}{64} \pi \alpha_s^3\;, & & c_1^B \!\! = \!\! \frac{27}{128} \pi \alpha_s^3 \;, & & d_0^B \!\! = \!\! \frac{1}{3} \pi^3 \alpha_s \; ; \\
& a_0^C \!\! = \!\! \frac{1}{112} \frac{\alpha_s^3}{\pi} \; , & & b_0^C \!\! = \!\! -\frac{1}{8} \pi \alpha_s^2 \;, & & b_1^C \!\! = \!\! \frac{3}{4} \pi \alpha_s^2 \; , \\
& c_0^C \!\! = \!\! \frac{79}{48}  \pi \alpha_s^3\;, & & c_1^C \!\! = \!\! \frac{845}{384} \pi \alpha_s^3 \;, & & d_0^C \!\! = \!\! 3 \pi^3 \alpha_s \; ; \\
& a_0^D \!\! = \!\! \frac{1}{1008} \frac{\alpha_s^3}{\pi} \; , & & b_0^D \!\! = \!\! \frac{23}{72} \pi \alpha_s^2 \;, & & b_1^D \!\! = \!\! \frac{1}{12} \pi \alpha_s^2 \; , \\
& c_0^D \!\! = \!\!  -\frac{47}{64} \pi \alpha_s^3\;, & & c_1^D \!\! = \!\! -\frac{1}{64} \pi \alpha_s^3 \;, & & d_0^D \!\! = \!\! \frac{1}{3} \pi^3 \alpha_s \; ,
\end{aligned}
\end{eqnarray}
where the ratios of $a^k_0$ to $b^k_1$ are equal for case $A$ to $D$. This implies that the mass curves of case $A$ to $D$ will be very similar, since if we neglect the $\langle g_s G^3 \rangle$ term which is much smaller than the $\langle \alpha_s G^2 \rangle$ term in Eq.(\ref{mass}), the mass of the oddball only depends on the ratio of $a^k_0$ to $b^k_1$.

For the $2^{+-}$ oddballs, the Wilson coefficients are
\begin{eqnarray}
\begin{aligned}
& a_0^A \!\! = \!\! - \frac{2}{81} \frac{\alpha_s^3}{\pi} \; ,  & & b_0^A \!\! = \!\! \frac{20}{3} \pi \alpha_s^2 \;, & & b_1^A \!\! = \!\! -\frac{20}{9} \pi \alpha_s^2 \; , \\
& c_0^A \!\! = \!\! \frac{205}{54} \pi \alpha_s^3\;, & & c_1^A \!\! = \!\! - \frac{40}{9} \pi \alpha_s^3 \;, & & d_0^A \!\! = \!\! \frac{20}{9} \pi^3 \alpha_s \; ; \\
& a_0^B \!\! = \!\! -\frac{1}{324} \frac{\alpha_s^3}{\pi} \; , & & b_0^B \!\! = \!\! \frac{5}{81} \pi \alpha_s^2 \;, & & b_1^B \!\! = \!\! \frac{10}{27} \pi \alpha_s^2 \; , \\
& c_0^B \!\! = \!\! \frac{415}{162} \pi \alpha_s^3\;, & & c_1^B \!\! = \!\! \frac{20}{27} \pi \alpha_s^3 \;, & & d_0^B \!\! = \!\! \frac{10}{9} \pi^3 \alpha_s \; ; \\
& a_0^C \!\! = \!\! - \frac{1}{324} \frac{\alpha_s^3}{\pi} \; , & & b_0^C \!\! = \!\! -\frac{115}{81} \pi \alpha_s^2 \;, & & b_1^C \!\! = \!\! \frac{10}{27} \pi \alpha_s^2 \; , \\
& c_0^C \!\! = \!\! - \frac{65}{162} \pi \alpha_s^3\;, & & c_1^C \!\! = \!\!  \frac{20}{27} \pi \alpha_s^3 \;, & & d_0^C \!\! = \!\! \frac{10}{9} \pi^3 \alpha_s \; ; \\
& a_0^D \!\! = \!\! - \frac{1}{324} \frac{\alpha_s^3}{\pi} \; , & & b_0^D \!\! = \!\! \frac{5}{81}  \pi \alpha_s^2 \;, & & b_1^D \!\! = \!\! \frac{10}{27}  \pi \alpha_s^2 \; , \\
& c_0^D \!\! = \!\! \frac{415}{162} \pi \alpha_s^3\;, & & c_1^D \!\! = \!\! \frac{20}{27} \pi \alpha_s^3 \;, & & d_0^D \!\! = \!\! \frac{10}{9}  \pi^3 \alpha_s \; ,
\end{aligned}
\end{eqnarray}
where $a^k_0$, $b^k_1$, and $c^k_1$ are equal for case $B$ to $D$. This implies that the mass curves of case $B$ to $D$ will be exactly equal, because they are determined by the Wilson coefficients $a^k_0$, $b^k_1$, and $c^k_1$.

To evaluate the oddball mass numerically, the following inputs are adopted \cite{Hao:2005hu}:
\begin{eqnarray}
\begin{aligned}
& \langle \alpha_s G^2\rangle = 0.06 \, \text{GeV}^4 \; ,\;
\langle g_s G^3\rangle = (0.27 \, \text{GeV}^2) \langle \alpha_s G^2\rangle \; ,\\
& ~~~~~~~~~~~\Lambda_{\overline{\text{MS}}} = 300 \, \text{MeV} \; ,\;
\alpha_s = \frac{-4\pi}{11 \ln (\tau \Lambda^2_{\overline{\text{MS}}})} \; ,
\end{aligned}
\end{eqnarray}
where the magnitude of the trigluon condensate, $\langle g_s G^3 \rangle$, is obtained from the dilute gas instanton model due to the lack of direct knowledge from experiment, while other parameters are commonly used in the literature.

In the QCDSR calculation, the parameter $\tau$ and the threshold $s_0$ are free parameters, proceeding from some requirements. Conventionally, two criteria are adopted in determining the $\tau$ \cite{Shifman, Reinders:1984sr, P.Col, Matheus:2006xi}.  First, the pole contribution (PC) should exceed that from the higher excited and continuum states. Therefore, one needs to evaluate the relative pole contribution over the total, the pole plus the higher excited and continuum states ($s_0 \to \infty$), for various $\tau$. In order to properly eliminate the contribution from higher excited and continuum states, the pole contribution is generally required to be more than $50 \%$. This criterion can be formulated as
\begin{eqnarray}
R^{k, \; \text{PC}}_{J} &=& \frac{L^{k}_{J^{PC} , \; 0}(\tau,s_0)}{L^{k}_{J^{PC} , \; 0}(\tau,\infty)} \; . \label{ratio-PC}
\end{eqnarray}
Second, the convergence of the OPE should be retained, that is the disregarded power corrections must be small. Namely, in the QCD side, the contribution of the leading condensate term should be smaller than $50 \%$ of the total contribution. For this aim, one needs to evaluate the relative weight of each term to the total on the QCD side. This criterion needs the following ratios
\begin{eqnarray}
R^{k, \; \text{G}^2}_{J} &=& \frac{\int_0^{s_0} e^{-s\tau} \mathrm{Im\Pi_{J^{PC}}^{k, \; \langle \alpha_s G^2\rangle}}(s) ds}{\int_0^{s_0} e^{-s\tau} \mathrm{Im\Pi_{J^{PC}}^{k, \; QCD}}(s) ds} \; , \label{ratio-GG} \\
R^{k, \; \text{G}^3}_{J} &=& \frac{\int_0^{s_0} e^{-s\tau} \mathrm{Im\Pi_{J^{PC}}^{k, \; \langle g_s G^3\rangle}}(s) ds}{\int_0^{s_0} e^{-s\tau} \mathrm{Im\Pi_{J^{PC}}^{k, \; QCD}}(s) ds} \; . \label{ratio-GGG}
\end{eqnarray}
Here, $k$ stands for cases $A$, $B$, $C$, and $D$, $\mathrm{Im\Pi_{J^{PC}}^{k, \langle \alpha_s G^2 \rangle}}(s)$ and $\mathrm{Im\Pi_{J^{PC}}^{k, \langle g_s G^3 \rangle}}(s)$ are the imaginary parts of the contributions from $\langle \alpha_s G^2 \rangle$ and $\langle g_s G^3 \rangle$, respectively.

To determine the characteristic value of $\sqrt{s_0}$, we carry out a similar analysis as in Refs.\cite{P.Col, Matheus:2006xi}. Therein, one needs to find out the proper value, which has an optimal window for the mass curve of the interested hadron. Within this window, the physical quantity, i.e., the mass of the oddball, is independent of the Borel parameter $\tau$ as much as possible. Through the above procedure one then obtains the central value of $\sqrt{s_0}$. However, in practice, it is normally acceptable to vary the $\sqrt{s_0}$ by about $0.2 \, \text{GeV}$ in the calculation of the QCDSR, which gives the lower and upper bounds and hence the uncertainties of $\sqrt{s_0}$.

With above preparation we numerically evaluate the mass spectra of the oddballs. For the $0^{+-}$ oddballs, we show the ratios $R_0^{A, PC}$ and $R_0^{A, G^3}$ as functions of Borel parameter $\tau$ in Fig.\ref{fig-1}(a) with different values of $\sqrt{s_0}$, 5.40, 5.60, and 5.80 GeV. We do not show the ratio $R_0^{A, G^2}$ in Fig.\ref{fig-1}(a), since it does not exist for the $0^{+-}$ oddballs. The dependency relations between oddball mass $M_{0^{+-}}^A$ and parameter $\tau$ are given in Fig.\ref{fig-1}(b). The parentheses in Fig.\ref{fig-1}(b) indicate the upper and lower limits of the valid Borel window for different values of $\sqrt{s_0}$. For the central value of $\sqrt{s_0}$, a smooth section, the so-called stable plateau, in $M_{0^{+-}}^A - \tau$ curve exists, suggesting the mass of the possible oddball. The situations for case $B$, $C$, and $D$ are shown in Figs.\ref{fig-2}, \ref{fig-3}, and \ref{fig-4}. We find that no matter what value the $\sqrt{s_0}$ takes, no optimal window for a stable plateau exists, where $M_{0^{+-}}^B$, $M_{0^{+-}}^C$ or $M_{0^{+-}}^D$ is nearly independent of the Borel parameter $\tau$. That means the current structures in Eqs.(\ref{current-0+-B}), (\ref{current-0+-C}), and (\ref{current-0+-D}) do not support the corresponding oddballs.

\begin{widetext}

\begin{figure}[hbpt]
\begin{center}
\includegraphics[width=6.0cm]{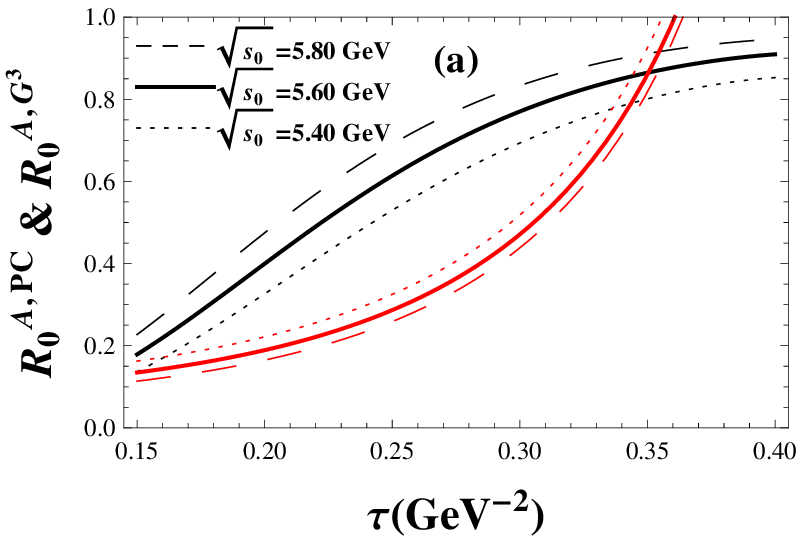}
\includegraphics[width=6.0cm]{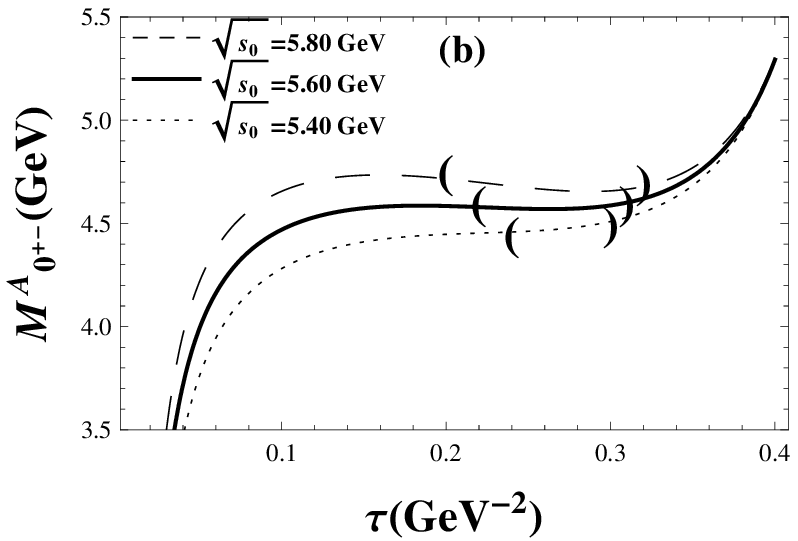}
\caption{(color). (a) The ratios $R_0^{A, PC}$ and $R_0^{A, G^3}$ in case $A$ as functions of the Borel parameter $\tau$ for different values of $\sqrt{s_0}$, where black lines represent $R_0^{A, PC}$ and red lines denote $R_0^{A, G^3}$. Note that the ratio $R_0^{A, G^2}$ is zero, so it does not exist in this figure. (b) The mass $M_{0^{+-}}^A$ as a function of the Borel parameter $\tau$ for different values of $\sqrt{s_0}$, where the parentheses indicate the upper and lower limits of the valid Borel window.}
\label{fig-1}
\end{center}
\end{figure}

\begin{figure}[hbpt]
\begin{center}
\includegraphics[width=6.0cm]{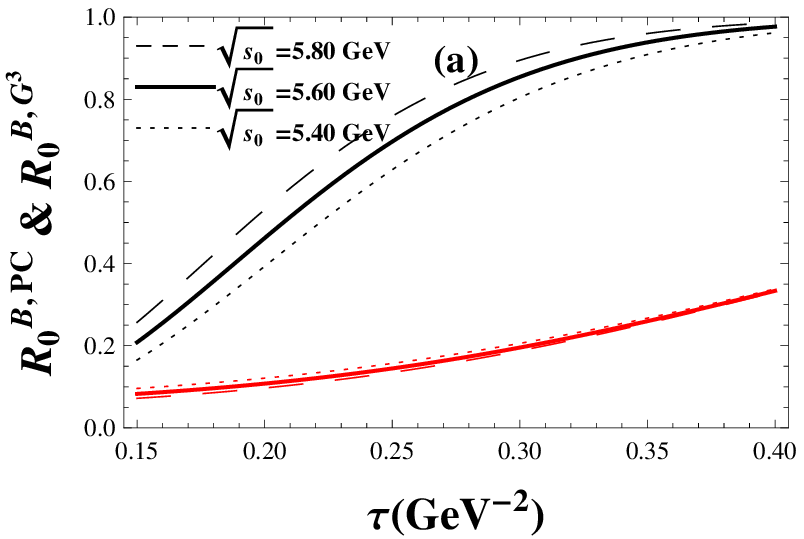}
\includegraphics[width=6.0cm]{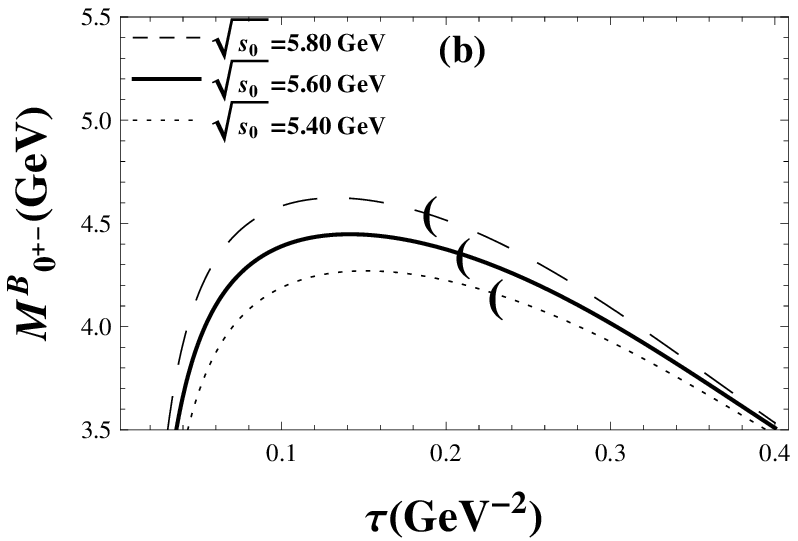}
\caption{(color). The same caption as in Fig.\ref{fig-1}, but for case $B$. Here the left parenthesis indicates the lower limit of the valid Borel window while the upper limit is out of the region. }
\label{fig-2}
\end{center}
\end{figure}

\begin{figure}[hbpt]
\begin{center}
\includegraphics[width=6.0cm]{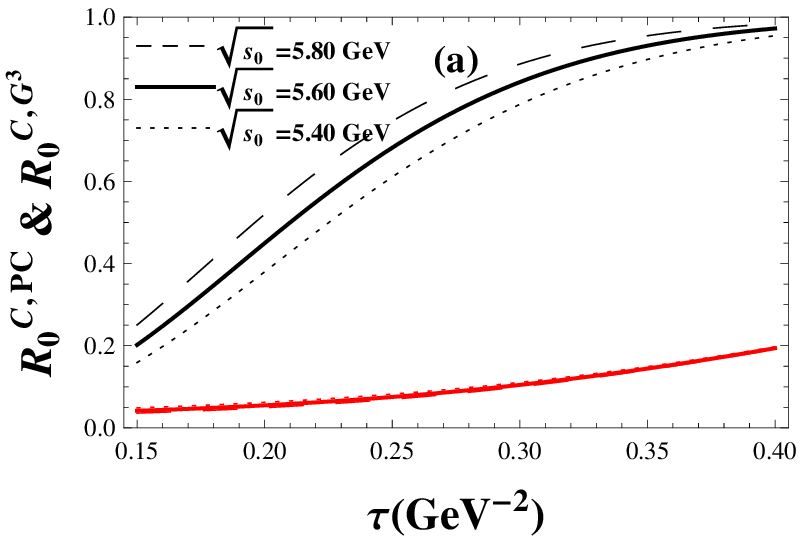}
\includegraphics[width=6.0cm]{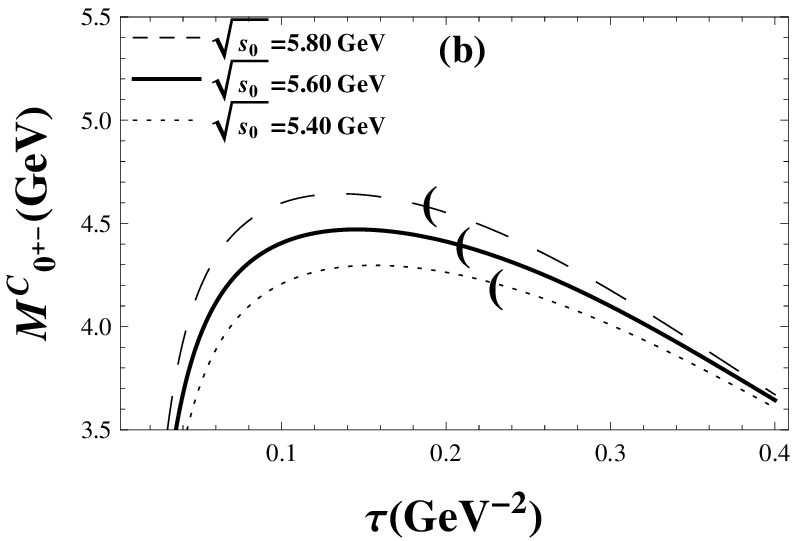}
\caption{(color). The same caption as in Fig.\ref{fig-2}, but for case $C$.}
\label{fig-3}
\end{center}
\end{figure}

\begin{figure}[hbpt]
\begin{center}
\includegraphics[width=6.0cm]{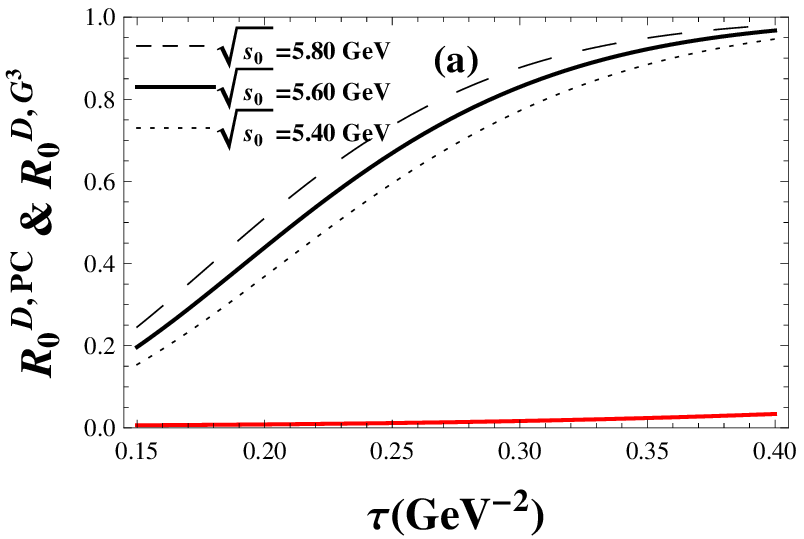}
\includegraphics[width=6.0cm]{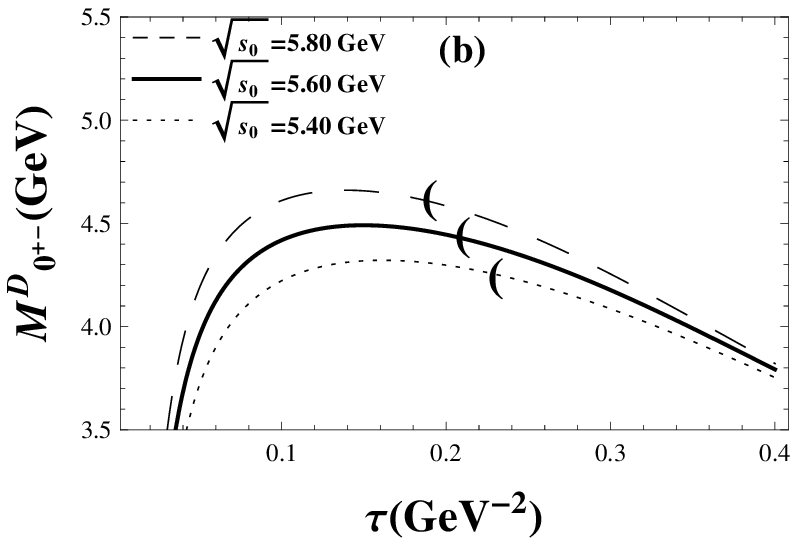}
\caption{(color). The same caption as in Fig.\ref{fig-2}, but for case $D$.}
\label{fig-4}
\end{center}
\end{figure}

\begin{figure}[hbpt]
\begin{center}
\includegraphics[width=6.0cm]{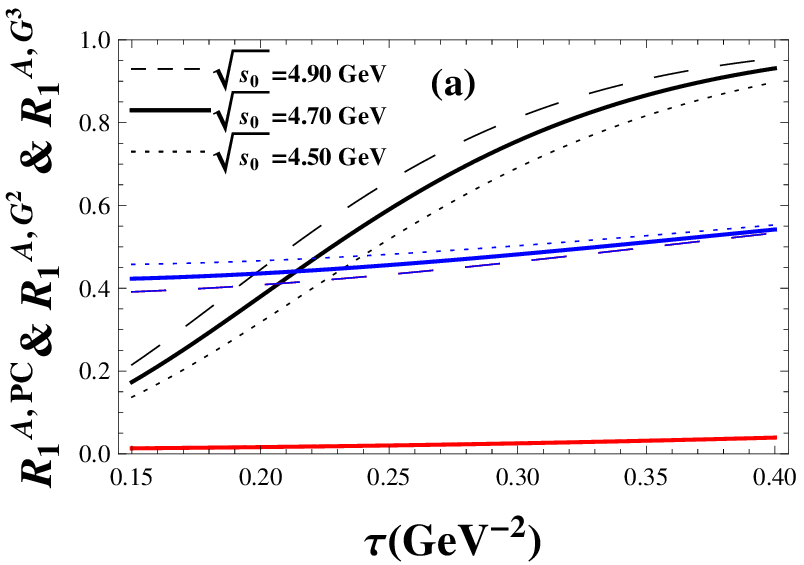}
\includegraphics[width=6.0cm]{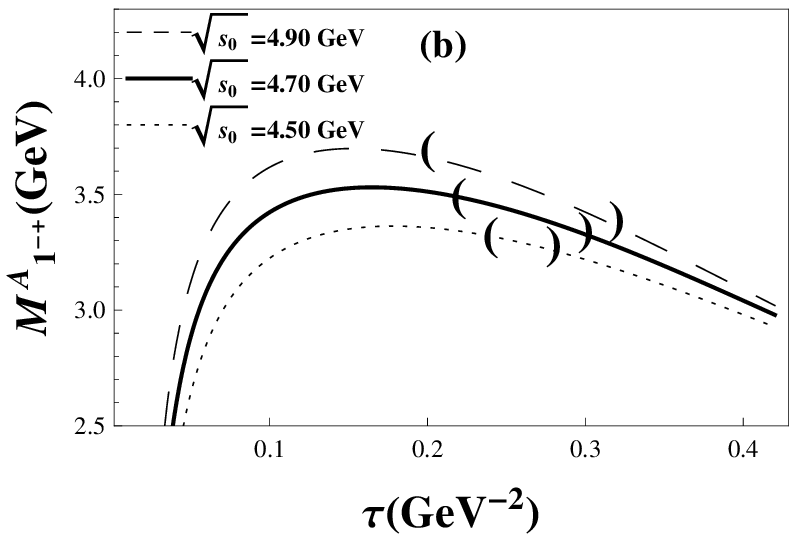}
\caption{(color). (a) The ratios $R_1^{A, PC}$, $R_1^{A, G^2}$, and $R_1^{A, G^3}$ in case $A$ as functions of the Borel parameter $\tau$ for different values of $\sqrt{s_0}$, where black lines represent $R_1^{A, PC}$, blue lines denote $R_1^{A, G^2}$, and red lines denote $R_1^{A, G^3}$. (b) The mass $M_{1^{-+}}^A$ as a function of the Borel parameter $\tau$ for different values of $\sqrt{s_0}$, where the parentheses indicate the upper and lower limits of the valid Borel window.}
\label{fig-5}
\end{center}
\end{figure}

\begin{figure}[hbpt]
\begin{center}
\includegraphics[width=6.0cm]{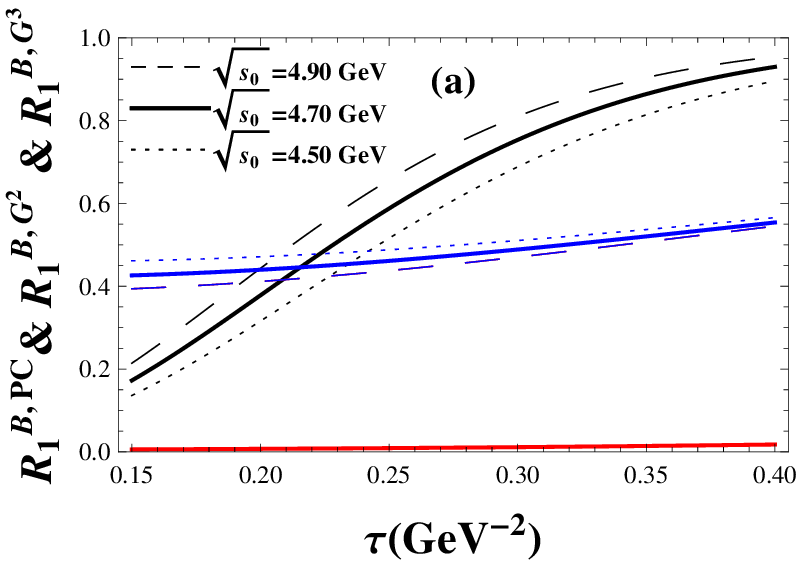}
\includegraphics[width=6.0cm]{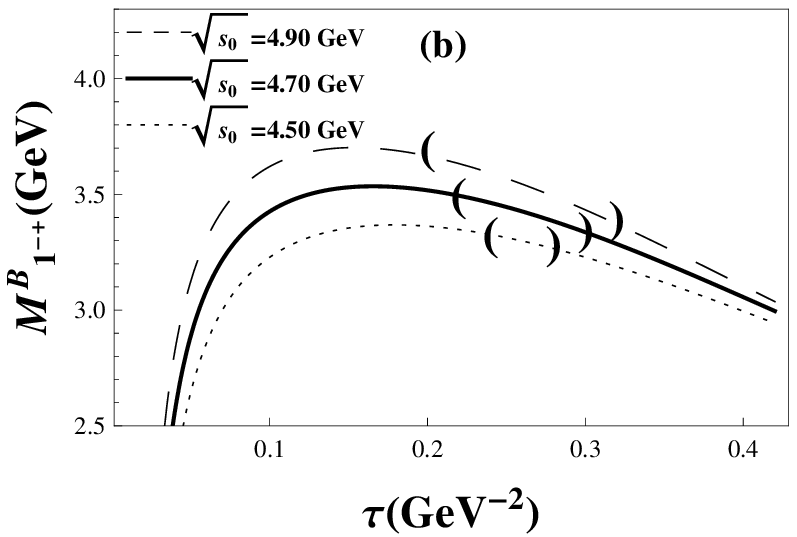}
\caption{(color). The same caption as in Fig.\ref{fig-5}, but for case $B$.}
\label{fig-6}
\end{center}
\end{figure}

\begin{figure}[hbpt]
\begin{center}
\includegraphics[width=6.0cm]{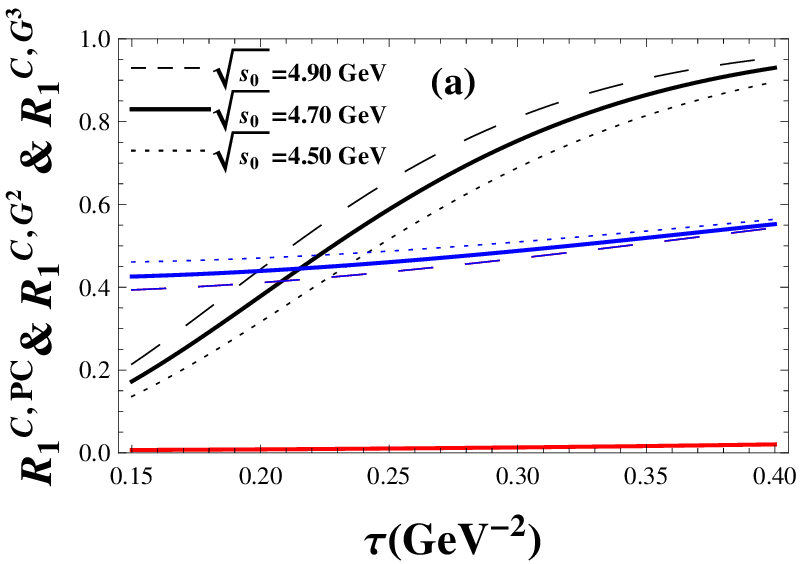}
\includegraphics[width=6.0cm]{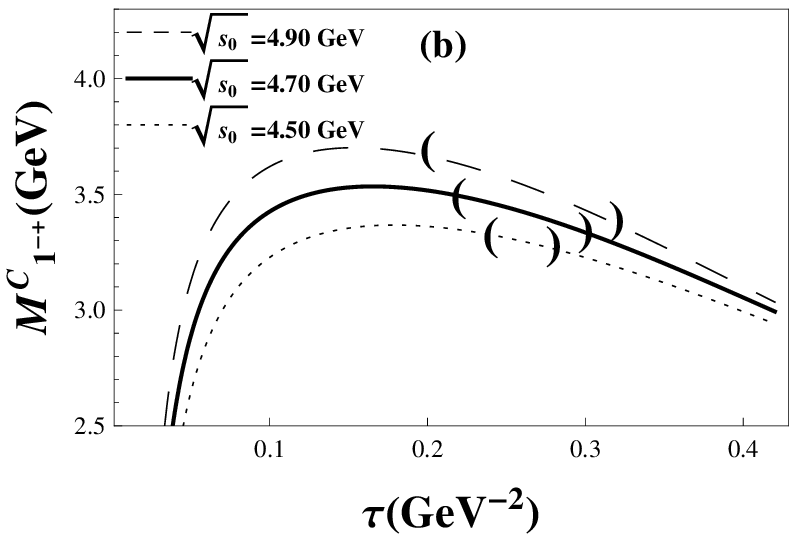}
\caption{(color). The same caption as in Fig.\ref{fig-5}, but for case $C$.}
\label{fig-7}
\end{center}
\end{figure}

\begin{figure}[hbpt]
\begin{center}
\includegraphics[width=6.0cm]{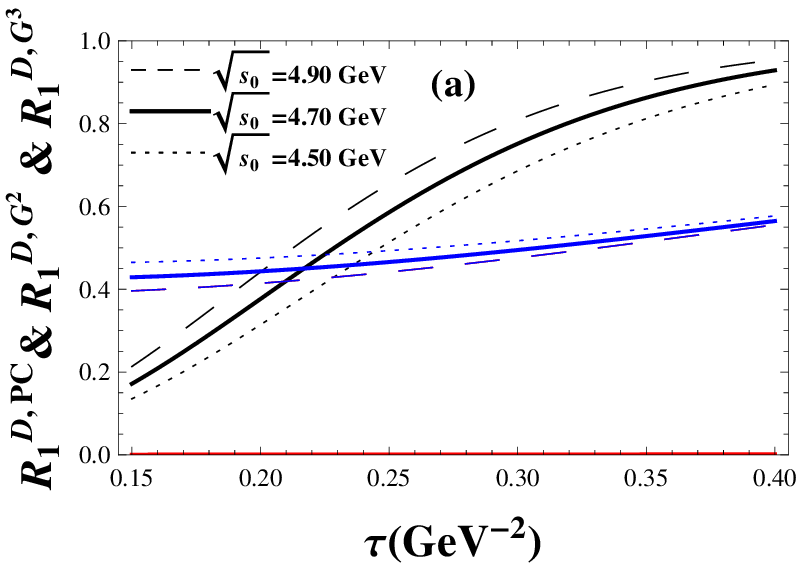}
\includegraphics[width=6.0cm]{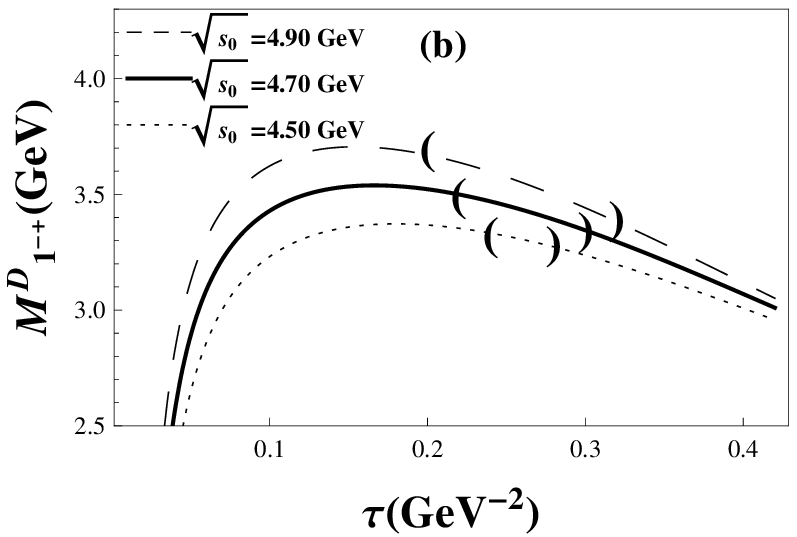}
\caption{(color). The same caption as in Fig.\ref{fig-5}, but for case $D$.}
\label{fig-8}
\end{center}
\end{figure}

\begin{figure}[hbpt]
\begin{center}
\includegraphics[width=6.0cm]{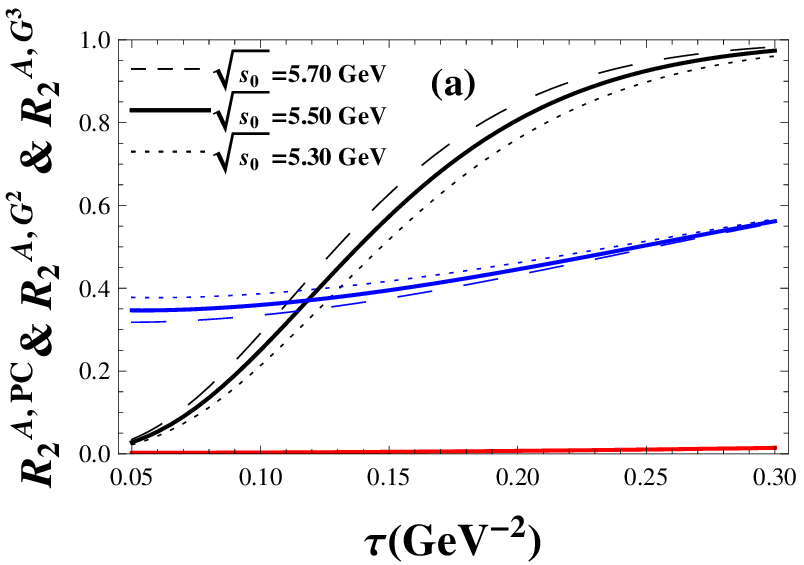}
\includegraphics[width=6.0cm]{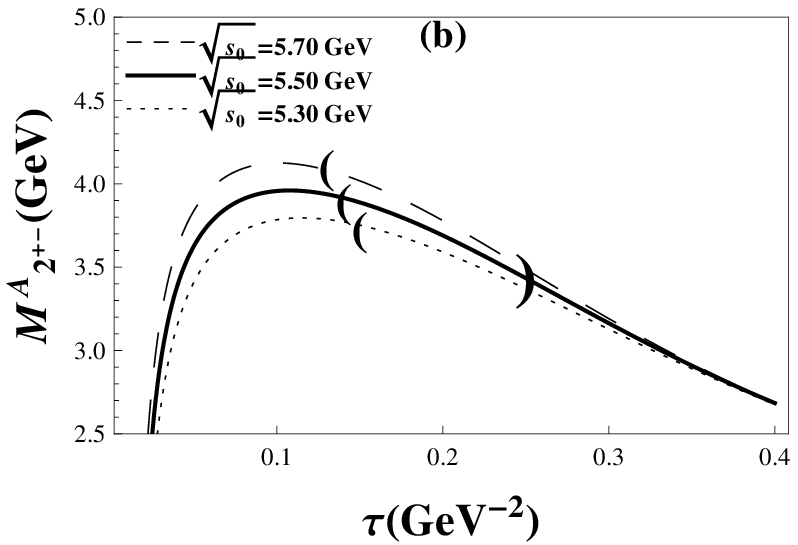}
\caption{(color). (a) The ratios $R_2^{A, PC}$, $R_2^{A, G^2}$, and $R_2^{A, G^3}$ in case $A$ as functions of the Borel parameter $\tau$ for different values of $\sqrt{s_0}$, where black lines represent $R_2^{A, PC}$, blue lines denote $R_2^{A, G^2}$, and red lines denote $R_2^{A, G^3}$. (b) The mass $M_{2^{+-}}^A$ as a function of the Borel parameter $\tau$ for different values of $\sqrt{s_0}$, where the parentheses indicate the upper and lower limits of the valid Borel window.}
\label{fig-9}
\end{center}
\end{figure}

\begin{figure}[hbpt]
\begin{center}
\includegraphics[width=6.0cm]{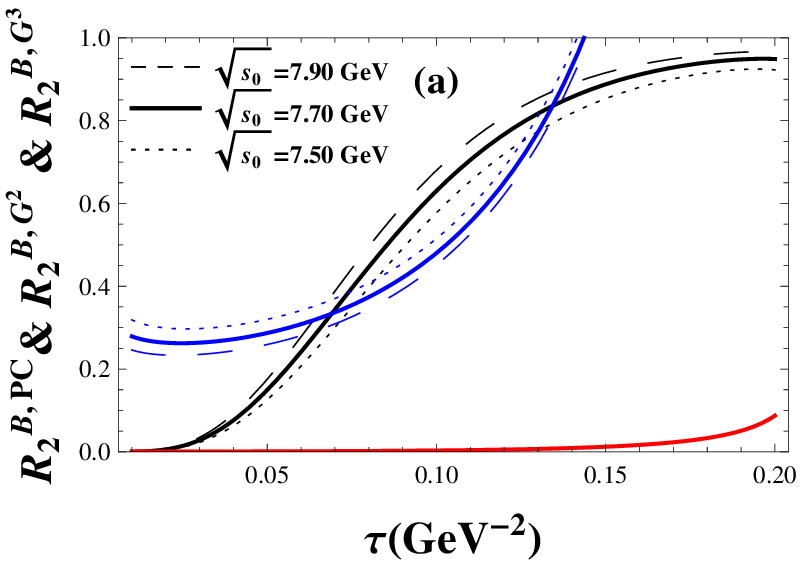}
\includegraphics[width=6.0cm]{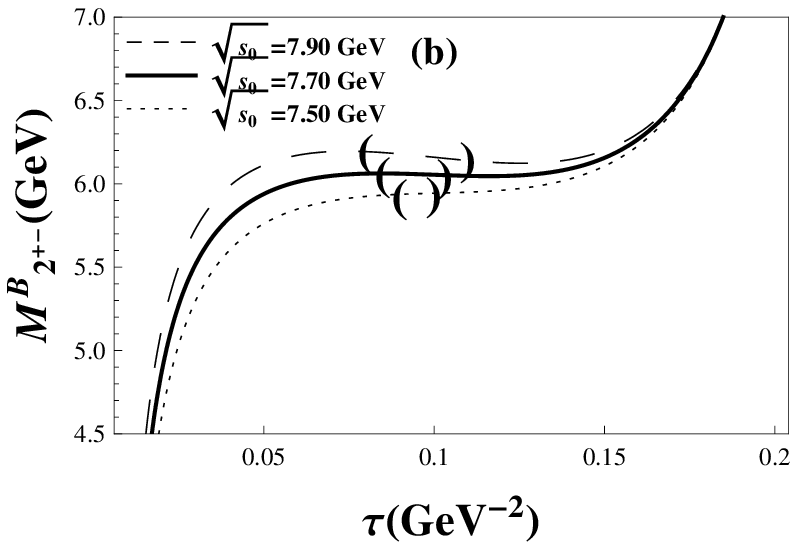}
\caption{(color). The same caption as in Fig.\ref{fig-9}, but for case $B$.}
\label{fig-10}
\end{center}
\end{figure}

\begin{figure}[hbpt]
\begin{center}
\includegraphics[width=6.0cm]{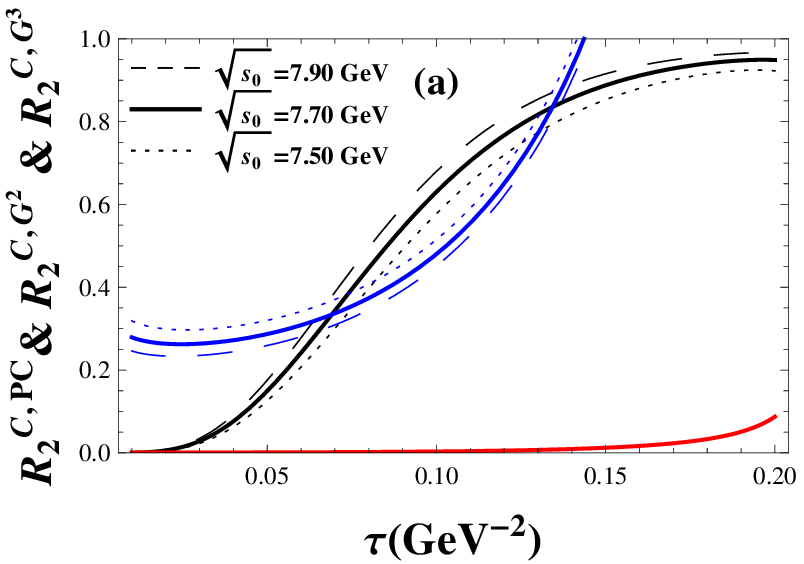}
\includegraphics[width=6.0cm]{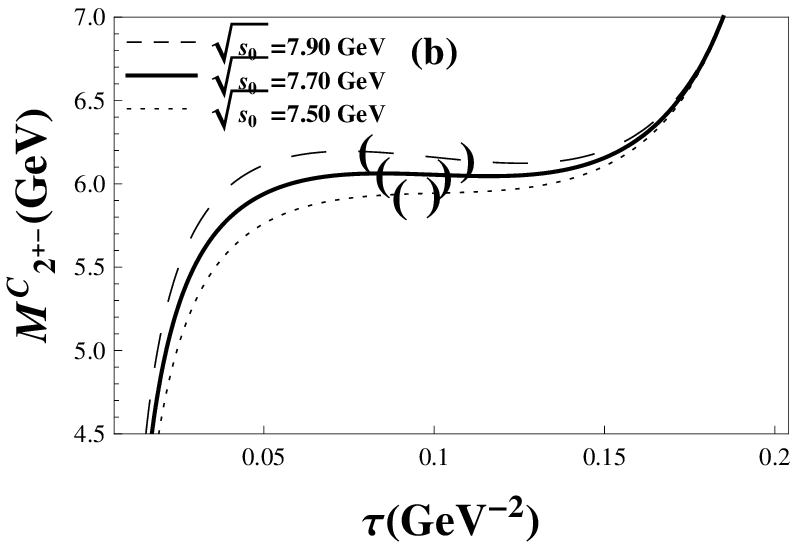}
\caption{(color). The same caption as in Fig.\ref{fig-9}, but for case $C$.}
\label{fig-11}
\end{center}
\end{figure}

\begin{figure}[hbpt]
\begin{center}
\includegraphics[width=6.0cm]{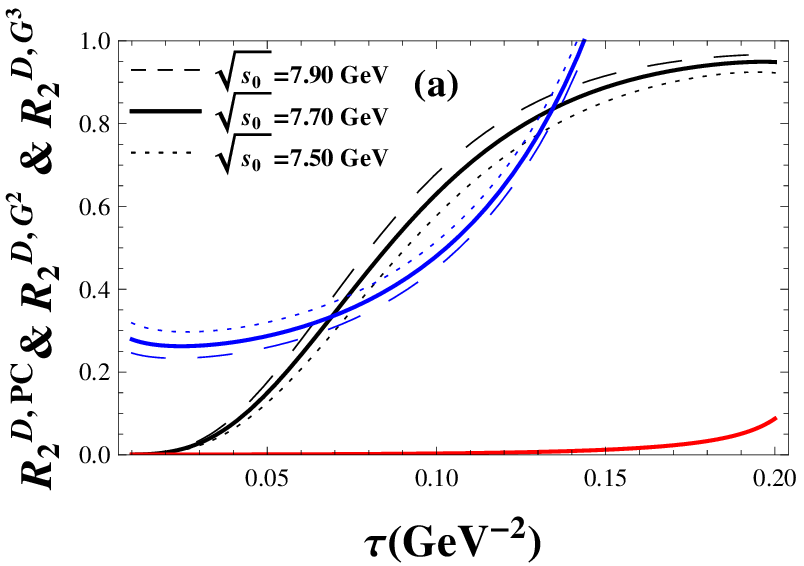}
\includegraphics[width=6.0cm]{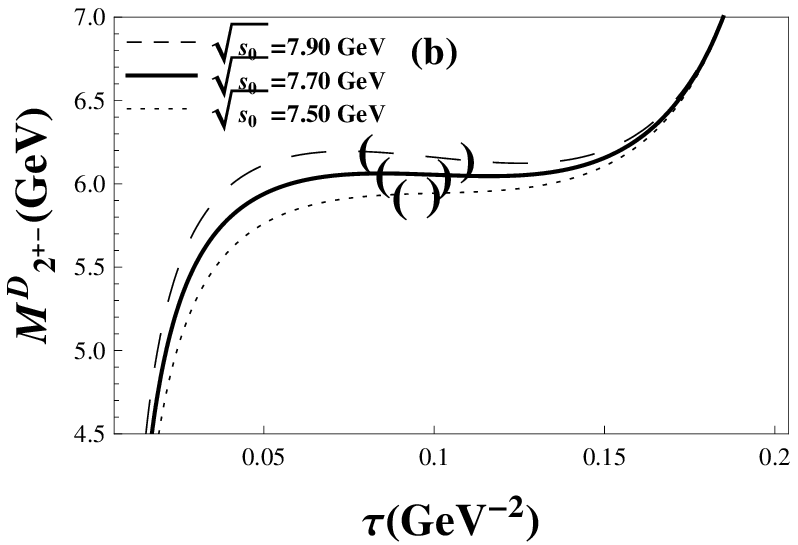}
\caption{(color). The same caption as in Fig.\ref{fig-9}, but for case $D$.}
\label{fig-12}
\end{center}
\end{figure}

\end{widetext}

\begin{table}[hbpt]
\caption{The lower and upper limits of the Borel parameter $\tau$ (GeV$^{-2}$) for $0^{+-}$, $1^{-+}$, and $2^{+-}$ oddballs for various cases with different $\sqrt{s_0}$ (GeV).}
\begin{center}
\renewcommand\arraystretch{1.1}
\begin{tabular}{|c|c|c||c|c|c||c|c|c||c|c|c|}
\hline \multicolumn{3}{|c||}  {$0^{+-}$ case $A$}  &  \multicolumn{3}{|c||}
{$0^{+-}$ case $B$} & \multicolumn{3}{|c||}  {$0^{+-}$ case $C$} & \multicolumn{3}{|c|}  {$0^{+-}$ case $D$}\\
\hline $\sqrt{s_0}$ & $\tau_{min}$  & $\tau_{max}$ & $\sqrt{s_0}$  & $\tau_{min}$ & $\tau_{max}$ & $\sqrt{s_0}$ & $\tau_{min}$  & $\tau_{max}$ & $\sqrt{s_0}$  & $\tau_{min}$ & $\tau_{max}$\\
\hline 5.80 & 0.20 & 0.32 & 5.80 & 0.19 & 0.50 & 5.80 & 0.19 & 0.60 & 5.80 & 0.19 & 1.20 \\
\hline 5.60 & 0.22 & 0.31 & 5.60 & 0.21 & 0.50 & 5.60 & 0.21 & 0.60 & 5.60 & 0.21 & 1.20 \\
\hline 5.40 & 0.24 & 0.30 & 5.40 & 0.23 & 0.50 & 5.40 & 0.23 & 0.60 & 5.40 & 0.23 & 1.20 \\
\hline
\hline \multicolumn{3}{|c||}  {$1^{-+}$ case $A$}  &  \multicolumn{3}{|c||}
{$1^{-+}$ case $B$} & \multicolumn{3}{|c||}  {$1^{-+}$ case $C$} & \multicolumn{3}{|c|}  {$1^{-+}$ case $D$}\\
\hline $\sqrt{s_0}$ & $\tau_{min}$  & $\tau_{max}$ & $\sqrt{s_0}$  & $\tau_{min}$ & $\tau_{max}$ & $\sqrt{s_0}$ & $\tau_{min}$  & $\tau_{max}$ & $\sqrt{s_0}$  & $\tau_{min}$ & $\tau_{max}$\\
\hline 4.90 & 0.20 & 0.32 & 4.90 & 0.20 & 0.32 & 4.90 & 0.20 & 0.32 & 4.90 & 0.20 & 0.32\\
\hline 4.70 & 0.22 & 0.30 & 4.70 & 0.22 & 0.30 & 4.70 & 0.22 & 0.30 & 4.70 & 0.22 & 0.30\\
\hline 4.50 & 0.24 & 0.28 & 4.50 & 0.24 & 0.28 & 4.50 & 0.24 & 0.28 & 4.50 & 0.24 & 0.28\\
\hline
\hline \multicolumn{3}{|c||}  {$2^{+-}$ case $A$}  &  \multicolumn{3}{|c||}
{$2^{+-}$ case $B$} & \multicolumn{3}{|c||}  {$2^{+-}$ case $C$} & \multicolumn{3}{|c|}  {$2^{+-}$ case $D$}\\
\hline $\sqrt{s_0}$ & $\tau_{min}$  & $\tau_{max}$ & $\sqrt{s_0}$  & $\tau_{min}$ & $\tau_{max}$ & $\sqrt{s_0}$ & $\tau_{min}$  & $\tau_{max}$ & $\sqrt{s_0}$  & $\tau_{min}$ & $\tau_{max}$\\
\hline 5.70 & 0.13 & 0.25 & 7.90 & 0.07 & 0.11 & 7.90 & 0.07 & 0.11 & 7.90 & 0.07 & 0.11\\
\hline 5.50 & 0.14 & 0.25 & 7.70 & 0.08 & 0.10 & 7.70 & 0.08 & 0.10 & 7.70 & 0.08 & 0.10\\
\hline 5.30 & 0.15 & 0.25 & 7.50 & 0.09 & 0.10 & 7.50 & 0.09 & 0.10 & 7.50 & 0.09 & 0.10\\
\hline
\end{tabular}
\end{center}
\label{window-3}
\end{table}

For the $1^{-+}$ oddballs, we show the corresponding figures in Figs.\ref{fig-5}-\ref{fig-8}. It should be noted that no matter what value the $\sqrt{s_0}$ takes, no optimal window for a stable plateau exists, where $M_{1^{-+}}^A$, $M_{1^{-+}}^B$, $M_{1^{-+}}^C$ or $M_{1^{-+}}^D$ is nearly independent of the Borel parameter $\tau$. That means the current structures in Eqs.(\ref{current-1-+A}), (\ref{current-1-+B}), (\ref{current-1-+C}), and (\ref{current-1-+D}) do not support the corresponding oddballs.

For the $2^{+-}$ oddballs, we show the corresponding figures in Figs.\ref{fig-9}-\ref{fig-12}. We notice that no matter what value the $\sqrt{s_0}$ takes, no optimal window for a stable plateau exists, where $M_{2^{+-}}^A$ is nearly independent of the Borel parameter $\tau$. That means the current structure in Eq.(\ref{current-2+-A}) does not support the corresponding oddball. However, for case $B$, the dependency relations between oddball mass $M_{2^{+-}}^B$ and parameter $\tau$ are given in Fig.\ref{fig-10}(b) with different values of $\sqrt{s_0}$, 7.50, 7.70, and 7.90 GeV. For the central value of $\sqrt{s_0}$ in Fig.\ref{fig-10}(b), a smooth section, the so-called stable plateau, in $M_{2^{+-}}^B - \tau$ curve exists, suggesting the mass of the possible oddball. The cases $C$ and $D$ have exactly the same mass curves as case $B$.

Our calculation shows that there possibly exists one $0^{+-}$ oddball and one $2^{+-}$ oddball, corresponding to the currents (\ref{current-0+-A}), (\ref{current-2+-B}), (\ref{current-2+-C}), and (\ref{current-2+-D}). That is
\begin{eqnarray}
\begin{aligned}
M_{0^{+-}}^A &= \!\!\! & 4.57 \pm 0.13 \, \text{GeV},   \\
M_{2^{+-}}^{B, \, C, \, D} &= \!\!\! & 6.06 \pm 0.13 \, \text{GeV},
\end{aligned}
\end{eqnarray}
where, the errors stem from the uncertainties of Borel parameter
$\tau$ and threshold parameter $\sqrt{s_0}$. From Fig.\ref{fig-1}(b), Fig.\ref{fig-10}(b), Fig.\ref{fig-11}(b), and Fig.\ref{fig-12}(b), it is obvious that these mass values of oddballs are quite stable and insensitive to the variation of $\tau$ and $\sqrt{s_0}$ within the proper windows of $\tau$. This is the main reason why our calculation yields small errors, similar as Refs.\cite{Huang:1998wj, Narison:1996fm} for instance. Hereafter, we refer the $0^{+-}$ oddball as $G_{0^{+-}}(4570)$, and $2^{+-}$ oddball as $G_{2^{+-}}(6060)$ in discussion.

In the literature, we notice that there existed some predictions of the unconventional quantum number oddballs in the lattice QCD calculation \cite{Morningstar:1999rf, Chen:2005mg, Gregory:2012hu} and the flux tube model \cite{Isgur:1984bm}. The comparison between their results and those in this paper are explicitly shown in Table.\ref{comparison}. Note that our result for the $0^{+-}$ oddball is larger than that in the flux tube model, where a mass of the $0^{+-}$ oddball was predicted to be about 2.79 GeV, whereas the lattice QCD calculation yielded even bigger results, 4.74, 4.78, and 5.45 GeV. A low-lying $1^{-+}$ oddball with mass of 1.68 GeV was estimated from the lattice QCD \cite{Ishikawa:1982kk, Carlson:1980kh}, however, flux tube model and the QCD Sum Rules calculations do not support it. In the $2^{+-}$ sector, the lattice QCD calculations give two close oddball masses, 4.14 and 4.23 GeV, which are much lower than the $G_{2^{+-}}(6060)$ predicted in this work.

\begin{widetext}
\begin{center}
\begin{table}[hbpt]
\caption{Comparison with Lattice QCD \cite{Morningstar:1999rf, Chen:2005mg, Gregory:2012hu, Ishikawa:1982kk}, and  the flux tube model \cite{Isgur:1984bm}, where a part of the unconventional quantum number oddballs with $J \leq 2$ were predicted. The notion ``X" denotes that there doesn't exist any oddball masses with this quantum number in the corresponding model.}
\vspace{0.3cm}
\renewcommand\arraystretch{1.1}
\begin{tabular}{|c|c|c|c|c|c|c|}
\hline
\hline $J^{PC}$  &  Lattice QCD \cite{Morningstar:1999rf} & Lattice QCD \cite{Chen:2005mg} &  Lattice QCD \cite{Gregory:2012hu} & Lattice QCD \cite{Ishikawa:1982kk} & Flux tube model \cite{Isgur:1984bm}  &  This work (QCD sum rules) \\
\hline $0^{+-}$ & 4.74 GeV & 4.78 GeV & 5.45 GeV & X & 2.79 GeV  & 4.57 GeV \\
\hline $1^{-+}$ & X & X & X & 1.68 GeV & X & X \\
\hline $2^{+-}$ & 4.14 GeV & 4.23 GeV & X & X &X & 6.06 GeV \\
\hline
\hline
\end{tabular}
\label{comparison}
\end{table}
\end{center}
\end{widetext}

\section{Production and decay analyses}

Experimentally, since the present measurement results for the glueball are either contradictory or at least non-conclusive, searching for clear evidence of glueball is now still an outstanding unsolved problem. This situation may be changed if measurement on unconventional glueballs makes progress. The oddballs of each unconventional quantum number are able to be detected in future experimental measurement due to their masses are attainable in most of the lepton colliders and the hadron colliders, such as the Belle, Super-B, and LHCb. Following we make a brief analysis on the feasibility of finding oddballs $G_{0^{+-}}(4570)$ and $G_{2^{+-}}(6060)$ in experiment.

The typical production modes of these lowest oddballs for each unconventional quantum number are exhibited in Table.\ref{production-mode}. All the parent particles in these processes are copiously produced in experiment, and hopefully decay to the oddballs with modest rates.

\begin{widetext}
\begin{center}
\begin{table}[h]
\caption{Typical production modes of the lowest oddballs for each unconventional quantum number.}
\vspace{0.3cm}
\renewcommand\arraystretch{1.1}
\begin{tabular}{|c|c|c|}
\hline
\hline $J^{PC}$  &  S-wave  & P-wave   \\
\hline   &   & $\Upsilon(1S) \to \bigg{\{} f_1(1285), \, \chi_{c1}, \, X(3872) \bigg{\}} + G_{0^{+-}}(4570)$  \\
$0^{+-}$ & $h_b \to \bigg{\{}f_1(1285), \, \chi_{c1}, \, X(3872) \bigg{\}} + G_{0^{+-}}(4570)$ & $\chi_{bJ} \to  \bigg{\{}\gamma, \, \omega, \, \phi, \, J/\psi, \, \psi(2S) \bigg{\}} + G_{0^{+-}}(4570)$  \\
 &  & $h_{b} \to \bigg{\{}\eta, \, \eta^\prime, \, \eta_c \bigg{\}} + G_{0^{+-}}(4570)$  \\
\hline  & $\Upsilon(1S) \to \eta_2(1645) + G_{2^{+-}}(6060)$ &  \\
$2^{+-}$ & $\chi_{b1, \, 2} \to \bigg{\{} h_1(1170), \,  h_c \bigg{\}} + G_{2^{+-}}(6060)$ & $\Upsilon(1S) \to  f_1(1285) + G_{2^{+-}}(6060)$ \\ & $h_b \to \bigg{\{} f_1(1285), \, f_2(1270), \, \chi_{c1, \, 2}, \bigg{\}} + G_{2^{+-}}(6060)$ & \\
\hline
\hline
\end{tabular}
\label{production-mode}
\end{table}
\end{center}

\begin{table}[h]
\caption{Typical decay modes of the lowest oddballs for each unconventional quantum number.}
\begin{center}
\renewcommand\arraystretch{1.1}
\begin{tabular}{|c|c|c|}
\hline
\hline $J^{PC}$  &  S-wave  & P-wave   \\
\hline   &   &  $G_{0^{+-}}(4570) \to \bigg{\{} \gamma, \, \omega, \, \phi, \, J/\psi \bigg{\}} + f_0(980)$ \\
$0^{+-}$ & $G_{0^{+-}}(4570) \to h_1(1170) +f_1(1285)$ & $G_{0^{+-}}(4570) \to h_1(1170) + \bigg{\{} \eta, \eta^\prime, \, \eta_c \bigg{\}}$ \\
 &  & $G_{0^{+-}} \to h_c + \bigg{\{} \eta, \eta^\prime \bigg{\}}$  \\
\hline $2^{+-}$ & $G_{2^{+-}}(6060) \to \bigg{\{} h_1(1170), \, h_c \bigg{\}} + f_1(1285)$ & $G_{2^{+-}}(6060) \to \bigg{\{}\gamma, \, \omega, \, \phi, J/\psi, \, \psi(2S) \bigg{\}} + f_1(1285)$ \\
\hline
\hline
\end{tabular}
\end{center}
\label{decay-mode}
\end{table}
\end{widetext}

To finally ascertain these oddballs, a straightforward procedures is to reconstruct them from its decay products, though the detailed characters of them need more work. Relatively, the exclusive processes are more transparent in this aim. Such typical decay modes of the lowest oddballs for each unconventional quantum number are shown in Table.\ref{decay-mode}.

These typical oddball production and decay processes are expected to be measurable in experiments. Detailed analysis on these oddballs production and decay issues is absent up to now. However, in the literature, many theoretical works \cite{Coyne:1980zd, Cheng:2006hu, Cotanch:2005ja, Zhao:2005nv} have analyzed the production and decay properties of the scalar ($0^{++}$) and tensor ($2^{++}$) glueballs. These investigations can shed light on the detailed analysis of the unconventional quantum number oddballs predicted by this work.

\section{Discussion and Conclusion}

In this work, by virtue of QCDSR we calculated the mass spectra of $0^{+-}$, $1^{-+}$, and $2^{+-}$ exotic glueballs. Note, though the unconventional quantum number oddballs will not mix with $q \bar{q}$ states, they can in principle mix with hybrids ($q \bar{q} g$) \cite{General:2006ed} and tetraquark states \cite{Xie:2013uha} with the same quantum number and similar mass, while naively the OZI suppression may hinder the mixing in certain degree \cite{Qiao:2014vva}. In this calculation the instanton and topological charge screening effects have not been taken into account, which as Forkel pointed out is important \cite{twogluon0++}, at least in cases like $0^{++}$ and $0^{-+}$ states. In this work, since the obtained results are very stable and the nonpertubative contributions are already quite large, we speculate the instantons contributions might be small.

According to the discussion in Ref.\cite{Latorre:1987wt}, the mixing occurs between two stable oddballs having the same quantum number and relatively small mass difference. Furthermore, it is notable that in QCD sum rules the relations of the currents with the resonances are built from the couplings. In some cases, a current does not yield a stable mass, which implies the coupling of the resonance to the current is possibly weak. In view of the above arguments, the mixing effect of resonances does not manifest in our calculation.

In conclusion, based on the interpolating currents with the unconventional quantum numbers of $J^{PC} = 0^{+-}$, $1^{-+}$, and $2^{+-}$, the oddball mass spectra are calculated in the framework of QCD sum rules. We find that one stable $0^{+-}$ oddball with mass of $4.57 \pm 0.13 \, \text{GeV}$ and one stable $2^{+-}$ oddball with mass of $6.06 \pm 0.13 \, \text{GeV}$ may exist, whereas, there is no stable $1^{-+}$ oddball found. We have briefly analysed these oddballs optimal production and decay mechanisms, which indicates that the long search elusive glueball is expected to be measured in BELLEII, Super-B, PANDA, and LHCb experiments.

%%%%%%%%%%%%%%%%%%%%%%%%%%%%%%%%%%%%%%%%%%%%%%%%%%%%%%%%%%%%%%%%%%%%%%%%%%%%
\acknowledgments

This work was supported in part by the Ministry of Science and Technology of the People's Republic of China (2015CB856703), and by the National Natural Science Foundation of China(NSFC) under the grants 11175249 and 11375200.

%%%%%%%%%%%%%%%%%%%%%%%%%%%%%%%%%%%%%%%%%%%%%%%%%%%%%%%%%%%%%%%%%%%%%%%%%%%%

%\appendix{\bf Appendix}


\begin{thebibliography}{9}

\bibitem{Gross:1973id}
  D.~J.~Gross and F.~Wilczek,
  Phys.\ Rev.\ Lett.\  {\bf 30}, 1343 (1973); H.~D.~Politzer, Phys.\ Rev.\ Lett.\  {\bf 30}, 1346 (1973).

\bibitem{Wilson:1974sk}
  K.~G.~Wilson,
  Phys.\ Rev.\ D {\bf 10}, 2445 (1974).

\bibitem{Morningstar:1999rf}
  C.~J.~Morningstar and M.~J.~Peardon,
  Phys.\ Rev.\ D {\bf 60}, 034509 (1999)
  [hep-lat/9901004].

\bibitem{Mathieu:2008me}
  V.~Mathieu, N.~Kochelev and V.~Vento,
  Int.\ J.\ Mod.\ Phys.\ E {\bf 18}, 1 (2009)
  [arXiv:0810.4453 [hep-ph]].

\bibitem{Chen:2005mg}
  Y.~Chen {\it et al.},
  Phys.\ Rev.\ D {\bf 73}, 014516 (2006)
  [hep-lat/0510074].

\bibitem{Gregory:2012hu}
  E.~Gregory, A.~Irving, B.~Lucini, C.~McNeile, A.~Rago, C.~Richards and E.~Rinaldi,
  JHEP {\bf 1210}, 170 (2012)
  [arXiv:1208.1858 [hep-lat]].

\bibitem{Ishikawa:1982kk}
  K.~Ishikawa, A.~Sato, G.~Schierholz and M.~Teper,
  Phys.\ Lett.\ B {\bf 120}, 387 (1983).

\bibitem{Isgur:1984bm}
  N.~Isgur and J.~E.~Paton,
  Phys.\ Rev.\ D {\bf 31}, 2910 (1985).

\bibitem{Jaffe:1975fd}
  R.~L.~Jaffe and K.~Johnson,
  Phys.\ Lett.\ B {\bf 60}, 201 (1976).

\bibitem{Chodos:1974je}
  A.~Chodos, R.~L.~Jaffe, K.~Johnson, C.~B.~Thorn and V.~F.~Weisskopf,
  Phys.\ Rev.\ D {\bf 9}, 3471 (1974).

\bibitem{Szczepaniak:1995cw}
  A.~Szczepaniak, E.~S.~Swanson, C.~R.~Ji and S.~R.~Cotanch,
  Phys.\ Rev.\ Lett.\  {\bf 76}, 2011 (1996) ; S.~R.~Cotanch, A.~P.~Szczepaniak, E.~S.~Swanson and C.~R.~Ji, Nucl.\ Phys.\ A {\bf 631}, 640C (1998); F.~J.~Llanes-Estrada, P.~Bicudo and S.~R.~Cotanch, Phys.\ Rev.\ Lett.\  {\bf 96}, 081601 (2006).

\bibitem{Colangelo:2007pt}
  P.~Colangelo, F.~De Fazio, F.~Jugeau and S.~Nicotri,
  Phys.\ Lett.\ B {\bf 652}, 73 (2007)
  [hep-ph/0703316].

\bibitem{Bellantuono:2015fia}
  L.~Bellantuono, P.~Colangelo and F.~Giannuzzi,
  arXiv:1507.07768 [hep-ph].

\bibitem{Brunner:2015yha}
  F.~Br¨¹nner and A.~Rebhan,
  Phys.\ Rev.\ Lett.\  {\bf 115}, no. 13, 131601 (2015)
  [arXiv:1504.05815 [hep-ph]].

\bibitem{Brunner:2015oqa}
  F.~Br¨¹nner, D.~Parganlija and A.~Rebhan,
  Phys.\ Rev.\ D {\bf 91}, no. 10, 106002 (2015)
  [arXiv:1501.07906 [hep-ph]].

\bibitem{Shifman}
  M.A. Shifman, A.I. Vainshtein and V.I. Zakharov,
  Nucl. Phys. {\bf B147}, 385 (1979); ibid, Nucl. Phys. {\bf B147},
  448 (1979).

\bibitem{Reinders:1984sr}
  L.~J.~Reinders, H.~Rubinstein and S.~Yazaki,
  Phys.\ Rept.\  {\bf 127}, 1 (1985).

\bibitem{twogluon0++}
  V.A. Novikov, M. A. Shifman, A.I. Vainshtein, and Valentin I.
  Zakharov, Nucl.\ Phys.\ B{\bf 165},\ 67(1980);
  M.A. Shifman,\ Z.\ Phys.\ C{\bf 9},\ 347(1981);
  E. Shuryak, Nucl.\ Phys.\ B{\bf 203},\ 116(1983);
  S. Narison,\ Z.\ Phys.\ C{\bf 26},\ 209(1984);
  M.A. Shifman, A.I. Vainshtein and V.I. Zakharov,
  Phys.\ Lett.\ {\bf B223}, 251(1989);
  S. Narison and G. Veneziano, Intern.\ J.\ Mod.\ Phys.\ {\bf A4},\
  2751(1989);
  E. Bagan and T.G. Steele, Phys.\ Lett.\ {\bf B243}, 413(1990);
  H. Forkel, Phys.\ Rev.\ D{\bf 64}, 034015(2001);
  H. Forkel, Phys.\ Rev.\ D{\bf 71}, 054008(2005).

\bibitem{Huang:1998wj}
  T.~Huang, H.~Y.~Jin and A.~L.~Zhang,
  Phys.\ Rev.\ D {\bf 59}, 034026 (1999)
  [hep-ph/9807391].

\bibitem{Narison:1996fm}
  S.~Narison,
  Nucl.\ Phys.\ B {\bf 509}, 312 (1998)
  [hep-ph/9612457].

\bibitem{twogluon0-+}
  E. Bagan and T.G. Steele, Phys.\ Lett.\ {\bf B243}, 413(1990);
  Ailin. Zhang and T. G. Steele, Nucl.\ Phys.\ A{\bf 728}, 165(2003).

\bibitem{Latorre:1987wt}
  J.~I.~Latorre, S.~Narison and S.~Paban,
  Phys.\ Lett.\ B {\bf 191}, 437 (1987).

\bibitem{Lu:1996tp}
  W.~-T.~Lu and J.~-P.~Liu,
  High Energy Phys.\ Nucl.\ Phys.\  {\bf 20}, 261 (1996).

\bibitem{Hao:2005hu}
  G.~Hao, C.~-F.~Qiao and A.~-L.~Zhang,
  Phys.\ Lett.\ B {\bf 642}, 53 (2006)  [hep-ph/0512214].

\bibitem{Yang:1950rg}
  C.~N.~Yang,
  Phys.\ Rev.\  {\bf 77}, 242 (1950).

\bibitem{Matheus:2006xi}
  R.~D'E.~Matheus, S.~Narison, M.~Nielsen and J.~M.~Richard,
  Phys.\ Rev.\ D {\bf 75}, 014005 (2007)
  [hep-ph/0608297].

\bibitem{Qiao:2014vva}
  C.~F.~Qiao and L.~Tang,
  Phys.\ Rev.\ Lett.\  {\bf 113}, 22, 221601 (2014)
  [arXiv:1408.3995 [hep-ph]].

\bibitem{YBDai}
  Y.~B.~Dai, C.~S.~Huang and H.~Y.~Jin,
  Z.\ Phys.\ C {\bf 60}, 527 (1993);
  Y.~B.~Dai, C.~S.~Huang, M.~Q.~Huang and C.~Liu,
  Phys.\ Lett.\ B {\bf 390}, 350 (1997);
  Y.~B.~Dai, C.~S.~Huang and M.~Q.~Huang,
  Phys.\ Rev.\ D {\bf 55}, 5719 (1997).

\bibitem{P.Col}
  P. Colangelo and A. Khodjamirian, in {\it At the frontier of
  particle physics / Handbook of QCD}, edited by M. Shifman (World
  Scientific, Singapore, 2001), arXiv:hep-ph/0010175.

\bibitem{Carlson:1980kh}
 C.~E.~Carlson, J.~J.~Coyne, P.~M.~Fishbane, F.~Gross and S.~Meshkov,
 Phys.\ Lett.\ B {\bf 99}, 353 (1981).

\bibitem{Coyne:1980zd}
  J.~J.~Coyne, P.~M.~Fishbane and S.~Meshkov,
  Phys.\ Lett.\ B {\bf 91}, 259 (1980).

\bibitem{Cheng:2006hu}
  H.~Y.~Cheng, C.~K.~Chua and K.~F.~Liu,
  Phys.\ Rev.\ D {\bf 74}, 094005 (2006);
  K.~T.~Chao, X.~G.~He and J.~P.~Ma,
  Eur.\ Phys.\ J.\ C {\bf 55}, 417 (2008).

\bibitem{Cotanch:2005ja}
  S.~R.~Cotanch and R.~A.~Williams,
  Phys.\ Lett.\ B {\bf 621}, 269 (2005);
  Y.~-B.~Yang {\it et al.}  [CLQCD Collaboration],
  Phys.\ Rev.\ Lett.\  {\bf 111}, no. 9, 091601 (2013).

\bibitem{Zhao:2005nv}
  Q.~Zhao and F.~E.~Close,
  Int.\ J.\ Mod.\ Phys.\ A {\bf 21}, 821 (2006);
  C.~D.~L¨¹, U.~G.~Meissner, W.~Wang and Q.~Zhao,
  Eur.\ Phys.\ J.\ A {\bf 49}, 58 (2013);
  R.~Zhu, JHEP {\bf 1509}, 166 (2015)
  [arXiv:1508.01445 [hep-ph]].

\bibitem{General:2006ed}
  I.~J.~General, S.~R.~Cotanch and F.~J.~Llanes-Estrada,
  Eur.\ Phys.\ J.\ C {\bf 51}, 347 (2007)
  [hep-ph/0609115].

\bibitem{Xie:2013uha}
  W.~Xie, L.~Q.~Mo, P.~Wang and S.~R.~Cotanch,
  Phys.\ Lett.\ B {\bf 725}, 148 (2013)
  [arXiv:1302.5737 [hep-ph]].


\end{thebibliography}
\end{document}